\documentclass[structabstract]{aa}   
\usepackage{graphicx}
\usepackage{natbib}
\usepackage[varg]{txfonts}
\usepackage{longtable}

\usepackage{txfonts}
\begin{document}
%
\title{Infrared/optical --- X-ray simultaneous observations \\
of X-ray flares in GRB~071112C and GRB~080506}



   \author{T.  Uehara\inst{1}  \and
	  M.  Uemura\inst{2} \and
	  K.  S. Kawabata\inst{2} \and\\
	  Y.  Fukazawa\inst{1} \and
	  R.  Yamazaki\inst{3} \and
	  A.  Arai\inst{4} \and
	  M.  Sasada\inst{1} \and
	  T.  Ohsugi\inst{2}  \and\\
	  T.  Mizuno\inst{1} \and
	  H.  Takahashi\inst{1}  \and
	  H.  Katagiri\inst{1} \and
	  T.  Yamashita\inst{5}  \and\\
	  M.  Ohno\inst{6} \and
	  G.  Sato\inst{6} \and
	  S.  Sato\inst{7}  \and
	  M.  Kino\inst{7} 
          }

   \institute{
         Department of Physical Science, Hiroshima University,
	 Kagamiyama 1-3-1, Higashi-Hiroshima 739-8526, Japan\\
	 \email{uehara@hep01.hepl.hiroshima-u.ac.jp}
        \and
	 Hiroshima Astrophysical Science Center, Hiroshima University, Kagamiyama
	 1-3-1, Higashi-Hiroshima 739-8526, Japan
        \and
        Department of Physics and Mathematics, 
         Aoyama Gakuin University, 5-10-1 Fuchinobe, 
         Sagamihara 252-5258, Japan 
        \and
         Faculty of Science, Kyoto Sangyo University, Motoyama, 
         Kamigamo, Kita-Ku, Kyoto-City 603-8555, Japan
	\and
	 National Astronomical Observatory of Japan, 2-21-1 Osawa, 
	 Mitaka, Tokyo 181-0015, Japan
	\and            
         Institute of Space and Astronautical Science, JAXA, 3-1-1
         Yoshinodai, Sagamihara, Kanagawa 229-8510, Japan
	\and
	 Department of Physics, Nagoya University, Furo-cho,
	 Chikusa-ku, Nagoya 464-8602, Japan
}

   \date{Received ; accepted }

 
  \abstract
   {}
   {We investigate the origin of short X-ray flares which are 
 occasionally observed in early stages of afterglows of 
 gamma-ray bursts (GRBs).
   }
   {We observed two events, GRB~071112C and GRB~080506, 
before the start of X-ray flares in the optical and near-infrared (NIR)
bands with the 1.5-m Kanata telescope.
In conjunction with published X-ray and optical data,
we analyzed densely sampled light curves of the early afterglows and 
spectral energy distributions (SEDs) 
in the NIR--X-ray ranges.
}
{
We found that the SEDs had a break between the optical and X-ray bands 
in the normal decay phases of both GRBs regardless
of the model for the correction of the interstellar extinction 
in host galaxies of GRBs.
   In the X-ray flares, X-ray flux increased by 3 and 15 times
in the case of GRB~071112C and 080506, respectively, 
and the X-ray spectra became harder than those in the normal decay phases.
No significant variation in the optical---NIR range was detected together 
with the X-ray flares.
}
   {These results suggest that the X-ray flares were associated with
either late internal shocks or external shocks from two-component jets.
}


   \maketitle
%

\section{Introduction}
Gamma-ray bursts (GRBs) are transient gamma-ray sources 
whose durations are $\sim$ 0.1--100~s \citep{Paciesas99}.  
Afterglows are occasionally observed after GRBs 
in a wide range of wavelengths from radio to X-rays 
(\citealt{Costa97, Paradijs97, Frail97}).  
The most plausible scenario for the origin of GRBs is
the internal shock model, 
in which the gamma-ray emission arises from a shock region generated 
by collisions between shells in a relativistic jet
(\citealt{pir99fireball, reviewZhang}).  
Afterglows are believed to be synchrotron emission from an 
external shock region between the shell and
interstellar medium (\citealt{Wang00, sar98grblc}).
It is widely acknowledged that a GRB having a long 
duration of $\sim$ 2.0--100~s, so-called `long GRBs', 
is related to a collapse of massive stars 
based on the fact that its direct connection with a type-Ic supernova
has been
confirmed (\citealt{Soderberg05, Della03, Hjorth03a, Stanek03, Malesani04}).

X-ray observations with the {\it Swift} satellite have discovered 
short flares in X-ray afterglows $10^{2-4}$~s after GRBs.  
These X-ray flares are unexpected in the framework of the standard 
external shock model because it predicts a monotonous decay of the 
X-ray afterglow with a power-law form \citep{Burrows05a}.
The X-ray flare is observed in a half of GRB afterglows \citep{Falcone07}.
The afterglow emission from the external shock could exhibit short-term 
modulations 
when a shell passes a high-density region of the interstellar medium  
\citep{Wang00}, or
slow shells catch up with the main shell \citep{Rees98}.
Such modulations of the emission from the external shock are candidates 
for the origin of the X-ray flare.
It is also possible that a late-time internal shock causes the X-ray 
flare (e.g. \citealt{Burrows05a, Zhang06, Chincarini07,
Butler_Kocevski07}).
This scenario with the late internal shock requires a long 
activity of the central engine of GRBs \citep{Ioka05}.

The optical and infrared observations of afterglows are important 
to evaluate the models for the X-ray flare 
because the external shock model predicts 
that an optical--infrared flare is associated with an X-ray flare.
No optical--infrared flare has been reported 
during the X-ray flares in previous observations, 
while a part of those observations were too sparse 
to investigate the detailed behavior 
of optical--infrared afterglows during the X-ray flares
(\citealt{Stanek07, Krimm07}).

Little is known about the temporal variation of spectral energy 
distributions (SEDs) associated with the X-ray flare.  This is 
because simultaneous multi-wavelength observations are required 
with a high time-resolution.
In addition, it is problematic 
to correct the interstellar extinction in host galaxies of GRBs
which is highly uncertain 
even if such a densely-sampled multi-wavelength data is available
(e.g. see \citealt{Stratta04, Kann06, Chen06extin,
Schady07, Starling07, Watson07}).

GRB~071112C and~080506 were detected by {\it Swift}/BAT
at 18:32:57.54~UT 12 November 2007 \citep{gcn7059} and 
17:46:21.22~UT 6 May 2008 \citep{gcn7685}, respectively.
X-ray flares were detected in both GRBs (\citealt{gcn7079, gcn7694}).
Here we report on our optical and NIR observations 
of those two GRBs using the Kanata 1.5-m telescope.  
Combined with X-ray spectra, these simultaneous optical and NIR 
data allowed us to investigate the interstellar extinction in 
the host galaxies, and thereby, variations of SEDs.
We describe the details of our observations in $\S$~2.  
We report on the temporal 
evolution of the optical and X-ray afterglows 
and SEDs in $\S$~3. 
In $\S$~4, we discuss the origin of the X-ray flare
using the internal and external shock models.  
Finally, we summarize our results in $\S$~5.

\subsection{Optical and NIR Observation}
\begin{table*}
  \caption{Kanata optical photometry of GRB~080506}\label{tab:log08}
  \begin{center}
    \begin{tabular}{ccccc}
     \hline \hline
     Filter & Time (s)$^*$ & Exposure (s) & mag & error \\
     \hline 
$R_{\it C}$ & 210.0 & 420.0 & 17.48 & $\pm$ 0.22 \\
$R_{\it C}$ & 264.0 & 528.0 & 17.59 & $\pm$ 0.27 \\
$R_{\it C}$ & 318.0 & 636.0 & 17.50 & $\pm$ 0.24 \\
$R_{\it C}$ & 375.0 & 750.0 & 17.81 & $\pm$ 0.33 \\
$R_{\it C}$ & 432.0 & 864.0 & 17.72 & $\pm$ 0.26 \\
$R_{\it C}$ & 496.0 & 992.0 & 17.82 & $\pm$ 0.28 \\
$R_{\it C}$ & 550.0 & 1100.0 & 17.90 & $\pm$ 0.36 \\
$R_{\it C}$ & 610.0 & 1220.0 & 17.73 & $\pm$ 0.26 \\
$R_{\it C}$ & 665.0 & 1330.0 & 17.89 & $\pm$ 0.30 \\
$R_{\it C}$ & 718.0 & 1436.0 & 18.24 & $\pm$ 0.39 \\
$R_{\it C}$ & 988.5 & 1977.0 & 18.14 & $\pm$ 0.13 \\
$R_{\it C}$ & 1138.5 & 2277.0 & 18.03 & $\pm$ 0.10 \\
$R_{\it C}$ & 1288.5 & 2577.0 & 17.98 & $\pm$ 0.10 \\
$R_{\it C}$ & 1439.5 & 2879.0 & 18.24 & $\pm$ 0.10 \\
$R_{\it C}$ & 1589.5 & 3179.0 & 18.14 & $\pm$ 0.10 \\
$R_{\it C}$ & 2820.5 & 5641.0 & 18.96 & $\pm$ 0.30 \\
$R_{\it C}$ & 2976.5 & 5953.0 & 18.69 & $\pm$ 0.24 \\
$R_{\it C}$ & 3134.5 & 6269 .0& 18.79 & $\pm$ 0.28 \\
$R_{\it C}$ & 3288.5 & 6577.0 & 18.88 & $\pm$ 0.33 \\
$R_{\it C}$ & 3519.0 & 7038.0 & 18.99 & $\pm$ 0.29 \\
$R_{\it C}$ & 3825.0 & 7650.0 & 19.27 & $\pm$ 0.35 \\
$R_{\it C}$ & 4136.0 & 8272.0 & 19.11 & $\pm$ 0.28 \\
$R_{\it C}$ & 4449.5 & 8899.0 & 19.22 & $\pm$ 0.37 \\
$R_{\it C}$ & 4830.2& 9660.3 & 19.19 & $\pm$ 0.26 \\
$R_{\it C}$ & 5371.2 & 10742.3 & 19.31 & $\pm$ 0.37 \\
$J$ & 319.8 & 639.6 & 16.30 & $\pm$0.08 \\
$J$ & 607.8 & 1215.6 & 16.64 & $\pm$0.13 \\
$J$ & 1063.5 & 2127.0 & 16.40 & $\pm$0.05 \\
$J$ & 1439.2 & 2878.3 & 16.62 & $\pm$0.06 \\
$J$ & 3132.3 & 6264.6 & 17.71 & $\pm$0.27 \\
$J$ & 3903.7 & 7807.4 & 18.14 & $\pm$0.39 \\
$J$ & 4677.9 & 9355.8 & 17.76 & $\pm$0.38 \\
$K_{\rm s}$ & 463.8 & 927.6 & 14.71 & $\pm$0.09 \\
$K_{\rm s}$ & 1288.9 & 2577.8 & 15.32 & $\pm$0.20 \\
$K_{\rm s}$ & 3904.6 & 7809.3 & 16.54 & $\pm$0.50 \\
     \hline
     \multicolumn{4}{l}{\footnotesize * Time since the GRB trigger}\\
    \end{tabular}
  \end{center}
\end{table*}

Our observations of GRB~071112C and GRB~080506 started at 18:36:21 (UT) 
12 November 2007 and 17:49:51 (UT) 6 May 2008, 
which were $\sim 324$~s and $\sim 210$~s 
after the GRB trigger times, respectively.  
Both observations were performed with TRISPEC attached to the Kanata 
1.5-m telescope at Higashi-Hiroshima Observatory of Hiroshima University.
TRISPEC is a simultaneous imager and spectrograph with polarimetry 
covering both optical and NIR wavelengths \citep{TRISPEC}.
We used the imaging mode of TRISPEC with the $V$, $J$, and $K_{\rm s}$  
band filters for GRB~071112C.  Instead of the $V$-band filter, 
we used the $R_{\it C}$-band filter for GRB~080506.
The observations continued for 4.5 and 5.7~ks, 
and we obtained 40 and 35 sets of three photometric band 
images for GRB~071112C and GRB~080506, respectively.
The central wavelength of
the TRISPEC's $R_{\it C}$ system is $\sim 620\;{\rm nm}$, 
slightly shifted from the standard one ($=645\;{\rm nm}$).  
The difference of the photometric systems is so small that 
we neglect it in our following discussion about spectral 
energy distributions.

We obtained differential magnitudes of the afterglows 
using a Java-based PSF photometry package
after making dark-subtracted and flat-fielded images.
We used a nearby field-star located at R.A.$=2^h36^m41^s.46$, 
Dec.$=28\degr 20\arcmin 53\arcsec .4$
as a comparison star for GRB~071112C.
The $V$, $J$, and $K_s$-band magnitudes 
of the comparison star were quoted from the Guide Star Catalog Version 2.3.2 
($V=14.43$) and 2MASS All-Sky Catalog of Point Sources ($J=13.238$ and 
$K_s=12.792$), respectively.
We checked systematic errors of
$V$, $J$, and $K_{\rm s}$ magnitudes
depending on comparison stars, and found that it is smaller than
0.28, 0.02, and 0.02~mag,
using neighboring stars.
\citet{gcn7094} contains differential magnitudes obtained 
by our optical observations of GRB~071112C.

For the differential photometry of GRB~080506, 
we used averages of magnitudes of
USNO B1.0 1289-0511223, 1289-0511261, 1289-0511235, 1289-0511157, 
and 1289-0511139 for the $R_{\it C}$ band,
2MASS 1289-0511197, 1289-0511261, 1289-0511235, 1289-0511157, 
and 1289-0511139 for the $J$ band,
and 2MASS 1289-0511197 and 1289-0511261 for the $K_{\rm s}$ band.
We checked systematic errors of $R_{\it C}$, $J$, and $K_{\rm s}$ magnitudes
depending on comparison stars, 
and found that it is smaller than 0.13, 0.01, and 0.06~mag,
using neighboring USNO B1.0 and 2MASS stars. 
Table~\ref{tab:log08} contains magnitudes obtained 
by our observations of GRB~080506.  
In this table, the magnitudes are averages in equally spaced bins 
in the logarithmic scale of time.  
The errors include both statistical and systematic ones.

\subsection{XRT analysis of GRB~071112C and GRB~080506}
XRT began observing GRB~071112C and GRB~080506 
at 2007 November 12 18:34:27 UT, i.e. at $T+90$~s and 
2008 May 6 17:48:47 UT, i.e. at $T+146$~s, respectively
($T$ represents a GRB trigger time).  
Both XRT data were processed using the HEASOFT package.
We extracted both source data with a 
rectangular 40$\arcsec \times 170\arcsec$
region for the Windowed Timing mode (WT),
and 40$\arcsec$ radius region for the Photon Counting mode (PC)
from the processed data.
Both GRB backgrounds were also extracted from
40$\arcsec \times 170\arcsec$ source of both ends region for the WT, 
and 192$\arcsec$ internal and 231$\arcsec$ outer annulus radius region
for the PC, far from the source.
Light curves were binned with a requirement of a minimum of 
30 photons per bin of WT and 20 photons per bin of PC for GRB~071112C,
while 30 photons per bin of WT and 40 photons per bin of PC for GRB~080506,

\section{Results}
\subsection{Optical and X-ray light curves}

\begin{figure*} 
 \centering
\includegraphics[width=125mm, angle=270]{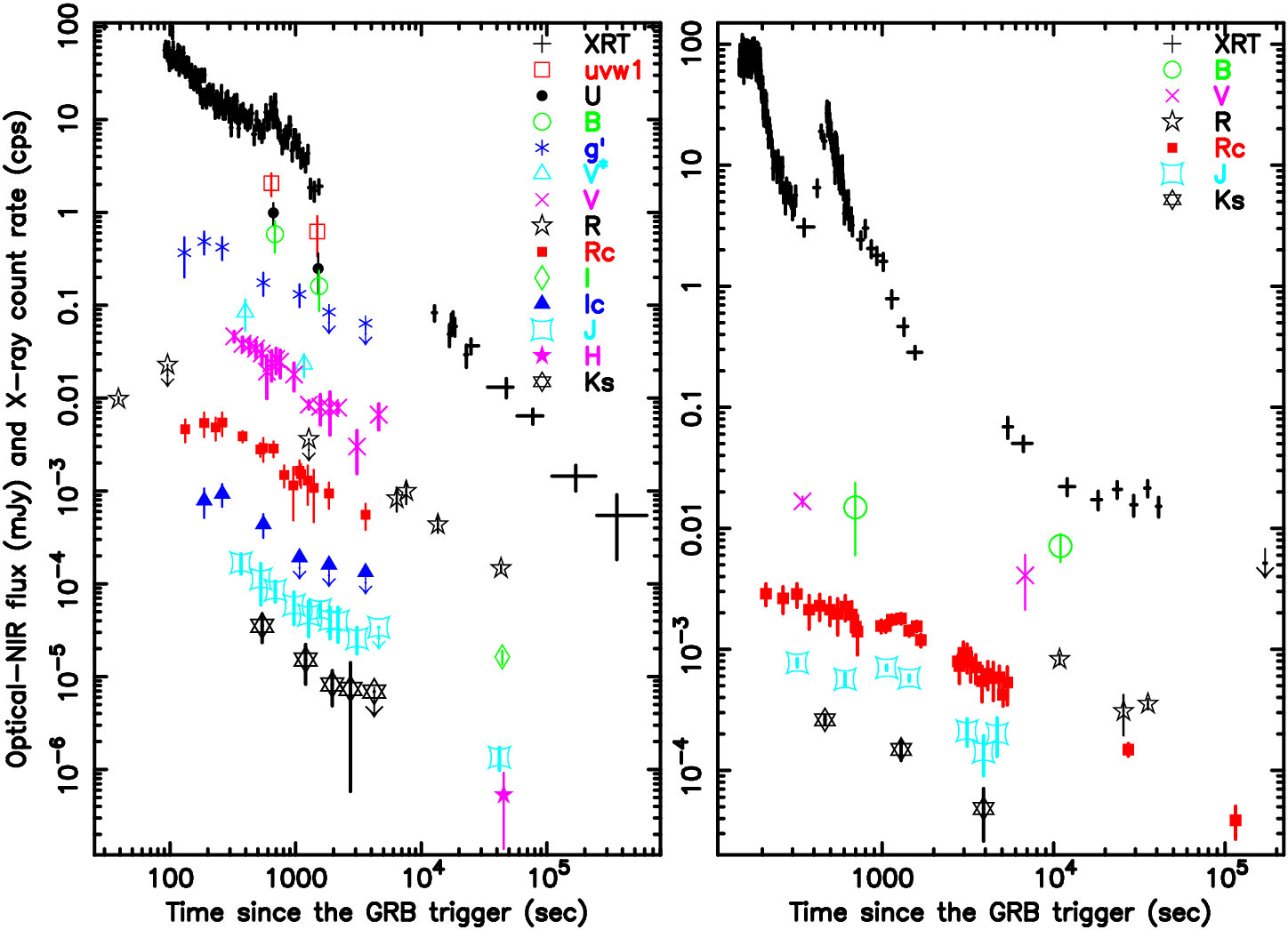}
  \caption{X-ray and NIR---UV light curves of 
  afterglows of GRB~071112C (left) and GRB~080506 (right).
  The flux density of the NIR---UV
  afterglows are shifted by $2$ in XRT, 
  $10^{0.5}$ in uvw1- \citep{gcn7080}, 
  $10^{0.0}$ in $U$- \citep{gcn7080}, 
  $10^{-0.5}$ in $B$- \citep{gcn7080}, 
  $10^{-1.0}$ in $g^\prime$- (\citealt{gcn7087, gcn7091}),
  $10^{-1.5}$ in $V^{*}$(UVOT)- \citep{gcn7080},
  $10^{-2.0}$ in $V$-, 
  $10^{-2.5}$ in $R$- (\citealt{gcn7065, gcn7066, gcn7067, gcn7089, gcn7135}), 
  $10^{-3.0}$ in $R_{\it C}$- (\citealt{gcn7078, gcn7087, gcn7091}), 
  $10^{-3.5}$ in $I$- \citep{gcn7135}, 
  $10^{-4.0}$ in $I_{\it C}$- (\citealt{gcn7087, gcn7091}), 
  $10^{-4.8}$ in $J$- \citep{gcn7135},
  $10^{-5.0}$ in $H$- \citep{gcn7135}, and 
  $10^{-5.5}$ in $K_{\rm s}$-band \citep{gcn7135} in the left panel.
  In the right panel, they are shifted by 
  $10^{-0.5}$ in $B$- \citep{gcn7693}, 
  $10^{-1.0}$ in $V$- \citep{gcn7693}, 
  $10^{-1.5}$ in $R$- (\citealt{gcn7689, gcn7690, gcn7700, gcn7709}), 
  $10^{-2.0}$ in $R_{\it C}$- (\citealt{gcn7688, gcn7696}), 
  $10^{-2.8}$ in $J$-, 
  $10^{-3.1}$ in $K_{\rm s}$-band.
   }
\label{fig:lc}
\end{figure*}

In figure~\ref{fig:lc}, 
we show the X-ray and NIR---UV light curves of 
afterglows of GRB~071112C (left) and GRB~080506 (right).
The X-ray observations by XRT are indicated by the crosses.  
The other symbols indicate the UV, optical, and NIR 
observations as described in the figure.
In addition to our observations by the Kanata telescope, this figure 
includes observations reported in GCN Circular:\citet{gcn7061}; 
\citet{gcn7062}; \citet{gcn7065}; \citet{gcn7066}; 
\citet{gcn7067}; \citet{gcn7069}; \citet{gcn7078}; \citet{gcn7080}; 
\citet{gcn7083}; \citet{gcn7084}; \citet{gcn7087}; \citet{gcn7089}; 
\citet{gcn7090}; \citet{gcn7091}; \citet{gcn7094}; \citet{gcn7092radio}; 
\citet{gcn7096radio}; \citet{gcn7135}; \citet{gcn7334}
for GRB~071112C 
and \citet{gcn7685}; \citet{gcn7686}; \citet{gcn7687};
\citet{gcn7688}; \citet{gcn7689}; \citet{gcn7690};
\citet{gcn7692}; \citet{gcn7693}; \citet{gcn7694}; \citet{gcn7696};
\citet{gcn7698}; \citet{gcn7700}; \citet{gcn7709}
for GRB~080506.
According to \citet{gcn7088z_host}, the host galaxy of GRB~071112C
is detected at $R_c = 24.5\pm 0.5$.
The host galaxy is so faint that its 
contribution to the afterglow is negligible in our analysis.
The host galaxy of GRB~080506 was not detected.

\begin{table*}
 \caption{Decay and spectral index of the X-ray afterglow of GRB~071112C}
\label{tab:grb071112c_x}
\centering          
\begin{tabular}{cccc}
\hline\hline
Region   & $T$ (s)  &$\alpha_{\rm X} \dag$ ($\chi^2$/d.o.f) & $\beta_{\rm X} \ddag $ \\ 
\hline
\it{Normal Decay}&100 $< T <$ 450, 1270 $< T <$ 1630 & $-$1.27 $\pm$0.05(104.8/96) & $-$0.83 $\pm$0.04    \\ 
\it{Flare Rise} & 450 $< T <$ 650                    & 2.51 $\pm$0.77 (7.3/6)     & $-$0.34$\pm$0.15  \\ 
\it{Flare Decay}& 650 $< T <$ 1270                   & $-$2.05 $\pm$0.26 (16.5/16)& $-$0.62 $\pm$0.06    \\ 
\it{Post-break}     & 12542 $< T <$ 147425             & $-$1.43 $\pm$0.04  (104.8/96) & $-$0.82 $_{-0.16}^{+0.08}$ \\ 
\hline
   \multicolumn{4}{@{}l@{}}{\hbox to 0pt{\parbox{85mm}{\footnotesize
       \par\noindent
        The uncertainties show the 90\% confidence levels of the parameters. 
       \par\noindent 
       $\dag$ X-ray decay index.
       \par\noindent
       $\ddag$ X-ray spectral index($\chi^2$/d.o.f = 587.7/298).
     }\hss}}
\end{tabular}
\end{table*}

The X-ray light curve of GRB~071112C
can be described with a broken power-law form;
$f \propto t^{\alpha_1} (t \leq t_{\rm break})$ and ,
$f \propto t^{\alpha_2} (t > t_{\rm break})$ 
with a weak X-ray flare between $T+450$~s and $T+1270$~s.
Excluding the X-ray flare phase, we calculated the decay indexes 
($\alpha_1$ and $\alpha_2$), and show them in table~\ref{tab:grb071112c_x}.  
The break time ($t_{\rm break}$) was estimated 
to be $7.6 \pm 0.7$~ks.  
Table ~\ref{tab:grb071112c_x} also includes the decay indexes during 
the rising and decay phases of the X-ray flare which we calculated 
assuming a single power-law form.
The optical light curves
show a brightening trend before $T+365$~s. 
After the maximum, the flux decayed as a single power-law
without a break as observed in the X-ray afterglow.
We confirmed that the decay indexes 
are same in all NIR to UV bands within errors.  
The decay index was, 
hence, calculated with all NIR---UV observations, 
and shown in table~\ref{tab:optical_para}.

The X-ray light curve of GRB~080506 (the right panel of figure~\ref{fig:lc}) 
is more complex than that of GRB~071112C.
Before $T+200$~s, the X-ray flux remained at a high level, 
which may be related to the prompt emission (\citealt{gcn7692, gcn7694}).
A steep decay was observed from $T+200$~s to $T+350$~s.
Between $T+350$~s and  $T+672$~s, 
it showed an X-ray flare centered at $T+$485 $\pm 2$~s. 
The X-ray afterglow finally entered a normal decay phase 
after the X-ray flare.
The optical and NIR light curves of GRB~080506 
can be described with a broken power-law model.
The power-law decay index changes after a break time at $T+1.3$~ks.
The calculated decay indexes are summarized in table~\ref{tab:optical_para} 
and \ref{tab:grb080506_x}.

\begin{table}
 \begin{center}
 \caption{Decay indexes of the optical light curves.}
 \label{tab:optical_para}
  \begin{tabular}{llll}
     \hline \hline
      GRB        & $T$ (s) &   $\alpha_O \dag$ & $\chi^2$ / d.o.f  \\ \hline
      GRB~071112C& 250 $< T <$ 10000   & $-$0.95($\pm$0.02)  & 31 / 44 \\ 
      GRB~080506 & 200 $< T <$ 1300    & $-$0.19($\pm$0.06)  & 59 / 76 \\ 
            & 1300 $< T < 2\times10^5$ & $-$0.83($\pm$0.03)  & \\ \hline
   \multicolumn{3}{@{}l@{}}{\hbox to 0pt{\parbox{85mm}{\footnotesize
       \par\noindent
       The uncertainties show the 90\% confidence levels of the parameters. 
       \par\noindent
       $\dag$ Optical decay index.
     }\hss}}
   \end{tabular}
 \end{center}
\end{table}
We succeeded in obtaining dense NIR, optical, and UV light 
curves around X-ray flares both in GRB~071112C and GRB~080506.
At the X-ray flare maxima, the X-ray flux increased by a factor of 
3.5 and 15 in GRB~071112C and GRB~080506, respectively.
On the other hand, the NIR---UV light curves exhibit 
no significant variations associated with the X-ray flares in both GRBs.  
The variation amplitudes are $<7\%$ in $V$ for GRB~071112C,
and $<14\%$ in $R_{\rm C}$ for GRB~080506.

\begin{table*}
 \caption{Decay and spectral index of the X-ray afterglow of GRB~080506}
\label{tab:grb080506_x}
\centering        
\begin{tabular}{cccc}
\hline\hline
Region & $T$ (s)& $\alpha_{\rm X}\dag$($\chi^2$/d.o.f) & $\beta_{\rm X} \ddag$  \\ 
\hline
\it{Prompt Emission} & 145-200 & prompt like                & $-$0.38 $\pm$0.03 \\ 
\it{Steep Decay}& 200-350     & $-$4.8 $\pm$0.2(43.3/30)    & $-$1.05 $\pm$0.06 \\ 
\it{Flare Rise} & 350-474     & 10.14 $\pm$0.72(23.8/6)     & $-$0.80 $\pm$0.18  \\
\it{Flare Peak} & 474-503     &      -                      & $-$1.02 $\pm$0.13  \\
\it{Flare Decay}& 503-672     & $-$6.41 $\pm$0.24(50.9/47)  & $-$1.49 $\pm$0.10  \\
\it{Normal Decay}  & 10900-70300  & $-$0.81 $\pm$0.05 (55.7/19) & $-$1.26 $_{-0.16}^{+0.17}$ \\ \hline
   \multicolumn{4}{@{}l@{}}{\hbox to 0pt{\parbox{85mm}{\footnotesize
       \par\noindent
       The uncertainties show the 90\% confidence levels of the parameters. 
       \par\noindent
       $\dag$ X-ray decay index.
       \par\noindent
       $\ddag$ X-ray spectral index($\chi^2$/d.o.f = 587.7/298).
     }\hss}}
\end{tabular}
\end{table*}


\subsection{XRT spectral analysis}
\label{sec:xrt_spec_ana}

\begin{figure*}
   \centering
   \includegraphics[width=120mm, angle=270]{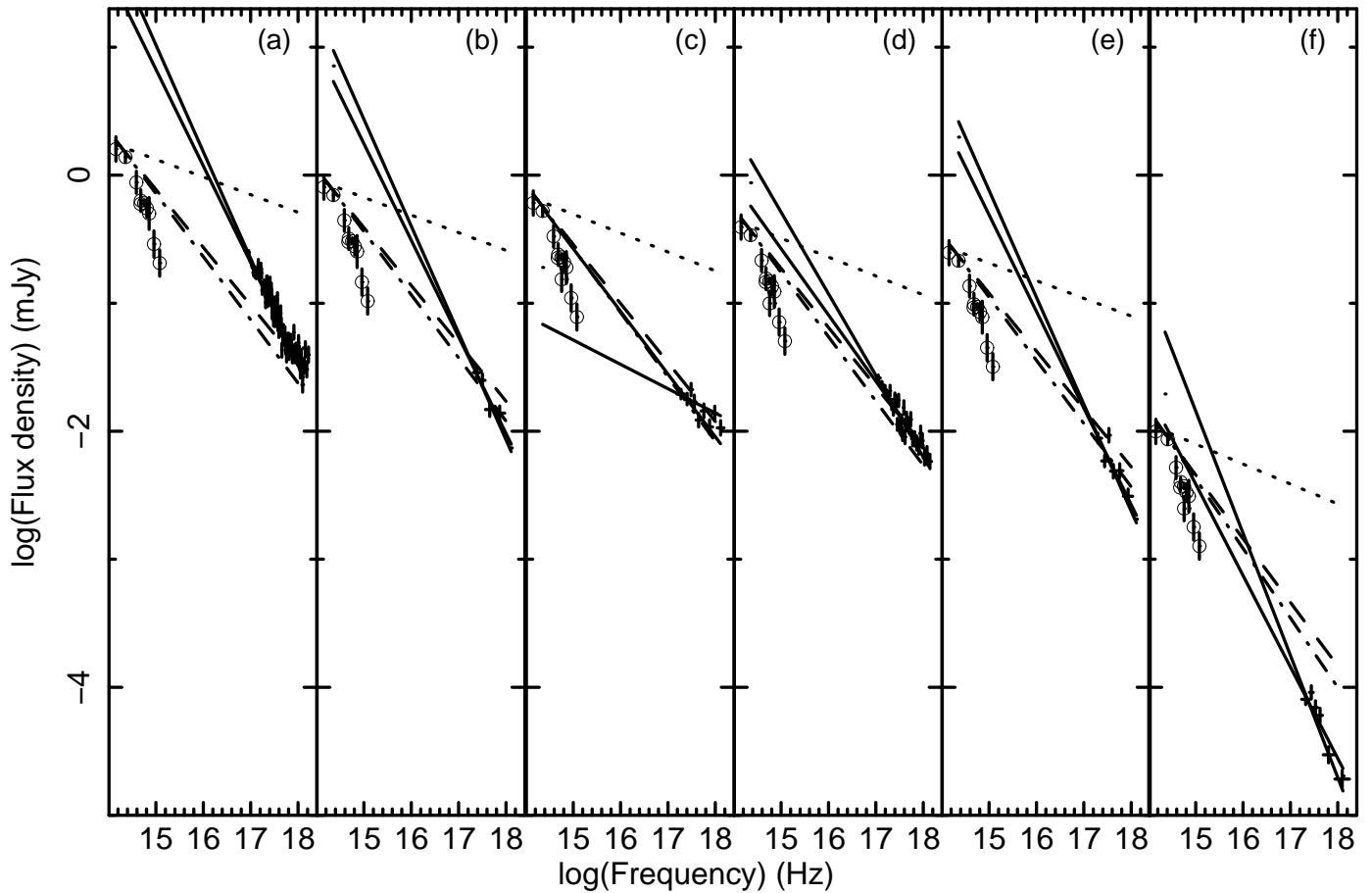}
  \caption{SEDs of the NIR---X-ray regime of GRB~071112C.
The 6 panels show the SED on 
(a) $T+100$ --- $T+365$~s, (b) $T+365$ --- $T+450$~s, 
(c) $T+450$ --- $T+650$~s, (d) $T+650$ --- $T+1270$~s, 
(e) $T+1270$ --- $T+1630$~s, and (f) $T+12542$ --- $T+147425$~s.
The crosses show the X-ray spectra.
The solid lines in the figure indicate 
the 90~\% fitting error region unabsorbed power-law model of the X-ray spectra. 
The open circles show observed infrared--UV SED. 
The doted, dashed, and dash-doted lines indicate
the best-fitted power-law models of the NIR---UV 
SEDs calculated 
with the extinction model in the MW,
LMC, and SMC, respectively.}
\label{fig:sed07}
\end{figure*}


In this subsection, we report the result of our analysis on X-ray 
spectra obtained with XRT, in particular, about the temporal variation
of the X-ray spectral index ($\beta_{\rm X}$).  
Based on the light curve analysis shown in the last section, 
we defined 4 and 6 phases for GRB~071112C 
and GRB~080506, as described in table~\ref{tab:grb071112c_x} 
and \ref{tab:grb080506_x}, respectively.  
The following analysis 
was performed for averaged X-ray spectra of each phase.
The average spectra of each phase were fitted 
with an absorbed power-law model with two absorption components.
The first component is the absorption in our galaxy.
According to \citet{Dickey90}, the galactic hydrogen column density
is $N_{\rm H}^{gal} = 7.4 \times 10^{20}$ and $1.66 \times 10^{21} {\rm cm}^{2}$ 
for the direction of GRB~071112C and GRB~080506, respectively.  
We used those $N_{\rm H}^{gal}$ to estimate 
the galactic absorptions for the GRBs.
The second component is the absorption in the host galaxy.  
The column density, $N_{\rm H}^{ext}$, was a free parameter in our 
analysis.


$N_{\rm H}^{ext}$ can be estimated from the absorption model, 
which depends on the redshift.
The redshift correction of the observed energy band is, hence, 
essential to correct the absorption.  
We performed the redshift correction to the energy band 
for GRB~071112C using a reported redshift of $z=0.82$ 
(\citealt{gcn7076z,gcn7086z}).
About GRB~080506, the redshift correction was not performed 
since the redshift is not determined.
We note that $\beta_{\rm X}$ is independent of the redshift 
correction, while $N_H^{\rm ext}$ depends on it.
Using the absorbed power-law model, we confirmed that the column 
density $N_{\rm H}^{ext}$ was constant in all phases for each GRB, 
while $\beta_{\rm X}$ changed.
We obtained the temporal variation of $\beta_{\rm X}$ in 0.3--8.0~keV band
by simultaneously fitting the spectra of all the phases by letting the
$N_{\rm H}^{ext}$ to be common among phases and $\beta_{\rm X}$ 
to be free for each phase.
The obtained $\beta_{\rm X}$ are summarized 
in table~\ref{tab:grb071112c_x} and \ref{tab:grb080506_x}, 
where the $\chi^2/dof$ of the fitting is 31/44 and 59/76 
for GRB~071112C and GRB~080506, respectively.
As shown in the table, the X-ray spectra in the rising phase of the
X-ray flares became harder than those in the normal decay phases
both in GRB~071112C and GRB~080506.
The column density $N_{\rm H}^{ext}$ of GRB~071112C 
was estimated to be $7.9\pm 3.9\times 10^{20}\;{\rm cm}^{-2}$.

\subsection{$\alpha$-$\beta$ relation in the X-ray afterglows}
\label{a-b_relation}


In the standard external shock model, the temporal decay 
index, $\alpha$, is related to the spectral slope, $\beta$
(e.g. \citealt{Zhang06}).
For example, $\alpha$ and $\beta$ in the X-ray regime have a relation of 
$\alpha_{\rm X} = 1.5 \beta_{\rm X}$ in the case that the X-ray 
band is between the synchrotron cooling frequency $\nu_{c}$ 
and the typical frequency $\nu_m$ under a homogeneous circum-burst medium.  
The $\alpha$ and $\beta$ during the normal decay phase 
of GRB~071112C satisfies this relation.
The electron energy distribution index, $p$, was estimated to be 
$p = 2.7$, inferred from $p = 1-\frac{4}{3}\alpha_{\rm X}$ 
and/or $p = 1-2\beta_{\rm X}$ (\citealt{Meszaros_Rees97, sar98grblc}).
This is similar to a typical $p$ in previously observed afterglows 
\citep{Liang07xrt}.
On the other hand, the $\alpha_{\rm X}$ and $\beta_{\rm X}$ in
GRB~080506 
do not follow the $\alpha$-$\beta$ relation.

\subsection{Spectral energy distribution of GRB~071112C}
\label{sec:sed07}
Figure \ref{fig:sed07} shows 
NIR---X-ray SEDs of GRB~071112C.  
The figure contains SEDs at 6 epochs, that is, 
(a) the rising phase of the optical afterglow ($T+$100 --- $T+$365~s), 
(b) the normal decay phase ($T+$36 --- $T$+450~s), 
(c) the rising phase of the X-ray flare ($T$+450 --- $T+$650~s), 
(d) the decay phase of the X-ray flare ($T+$650 --- $T+$1270~s), 
(e) the normal decay phase ($T+$1270 --- $T+$1630~s), and 
(f) the post-break phase ($T+$12542 --- $T+$147425~s). 
The crosses show the X-ray spectrum.
The solid lines indicate the 90\% confidence region of unabsorbed 
power-law model of the X-ray spectra. 
The open circles show the observed NIR---UV fluxes.
In all phases except for the rising phase of X-ray flare, 
the NIR---UV fluxes are much lower than those expected 
from the extrapolated power-law components of the X-ray spectra.
Thus, the NIR---UV fluxes are definitely reduced by the 
interstellar extinction.

We estimated the extinction in the $V$-band, $A_V$, from $N_{\rm H}$ 
obtained from the X-ray spectral analysis.  
The ratio, $N_{\rm H}/A_V$, has been reported in our and nearby
galaxies, that is, 
$N_{\rm H}/A_V=1.6\times 10^{21}\; {\rm cm}^{-2}$ for Milky Way (MW), 
$7.6\times 10^{21}\; {\rm cm}^{-2}$ for the Large Magellanic Cloud (LMC),
and $1.5\times 10^{22}\; {\rm cm}^{-2}$ for the Small Magellanic Cloud (SMC)
(\citealt{Pei92dust, Weingartner_Draine00}).  
Using those 3 models, 
we corrected the extinction in the $V$-band.  
The conversion from $A_V$ to $A_{\lambda}$ of 
the other bands was performed based on \citet{car89extinct}.  
In figure~\ref{fig:sed07}, 
the dotted, dashed, and dash-dotted lines indicate the 
best-fitted power-law models for the corrected NIR---UV SEDs with the 
MW, LMC, and SMC models, respectively.
In the case that the NIR---UV extinction was corrected 
with the LMC or SMC model,
the X-ray flux has an excess over the SED extrapolated 
from the NIR---UV regime 
in panel~(a) of figure~\ref{fig:sed07}.
Such an X-ray excess is inconsistent with the standard external
shock model.
In the case that the extinction model is the MW model,
the SED should have a break between the optical and UV band
in panel~(f),
while such a break is not clearly seen in the observed 
NIR---UV SED.
However, the break could be in UV--X-ray SED,
when considering the error of X-ray spectral index.

In the framework of the GRB external shock model, 
 NIR---X-ray SEDs can be described with a single power-law model 
 if the cooling frequency, $\nu_c$, is below the NIR band.  
 Alternatively, SEDs have a break in the case that $\nu_c$ lies 
 between the NIR---X-ray bands.   
 As mentioned above and shown in figure~\ref{fig:sed07}, 
 a single power-law model cannot reproduce the observed SED of 
 the NIR--X-ray bands regardless of the ambiguity of the 
 extinction.  Therefore, the standard external shock model can 
 explain the observed SED only when $\nu_c$ lies between the 
 NIR---X-ray bands.  Then, the SED should have a break between the 
 NIR---X-ray bands.  This break actually appears in the case that 
 $N_{\rm H}^{ext}/A_V$ in the host galaxy of GRB~071112C is 
 between the values of the MW and LMC models.

\subsection{Spectral energy distribution of GRB~080506}
\label{sec:sed08}
\begin{figure*}
   \centering
   \includegraphics[width=120mm,  angle=270]{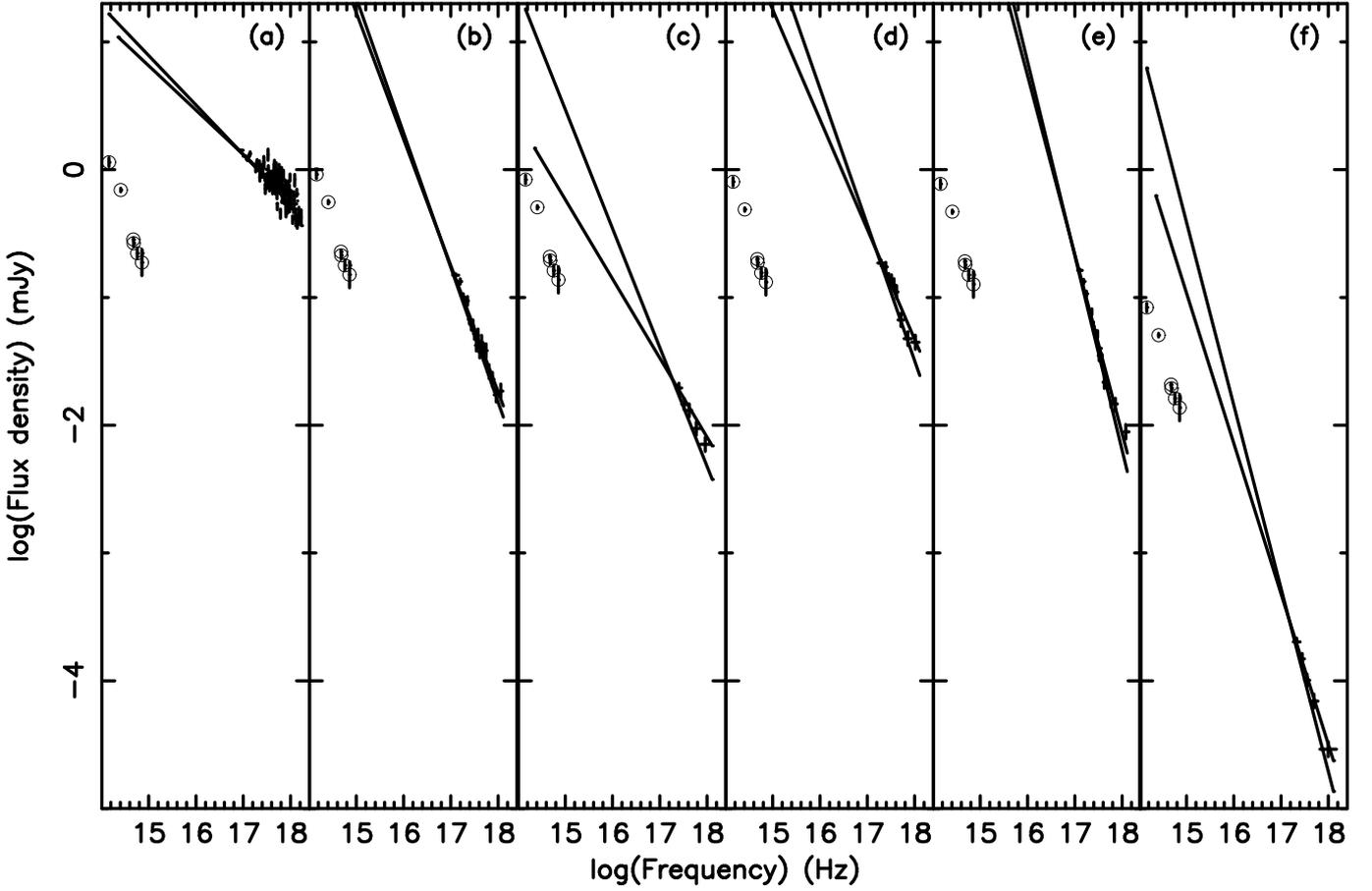}
\caption{SEDs of the NIR---X-ray regime of GRB~080506.
These epochs correspond to the phases defined in table~4, 
 that is, (a) Prompt Emission on $T+145$ --- $T+200$~s, 
 (b) Steep Decay on $T+200$ --- $T+350$~s, 
 (c) Flare Rise on $T+350$ --- $T+474$~s,
 (d) Flare Peak on $T+474$ --- $T+503$~s, 
 (e) Flare Decay on $T+503$ ---- $T+672$~s, and 
 (f) Normal Decay on $T+10900$ --- $T+70300$~s. 
The solid liens show 
90~\% fitting error region of X-ray spectra that corrected for interstellar extinction. 
No redshift
correction was performed to the energy band.}
\label{fig:sed08}
\end{figure*}

\begin{figure*}
  \begin{center}
   \centering
   \includegraphics[width=70mm,  angle=270]{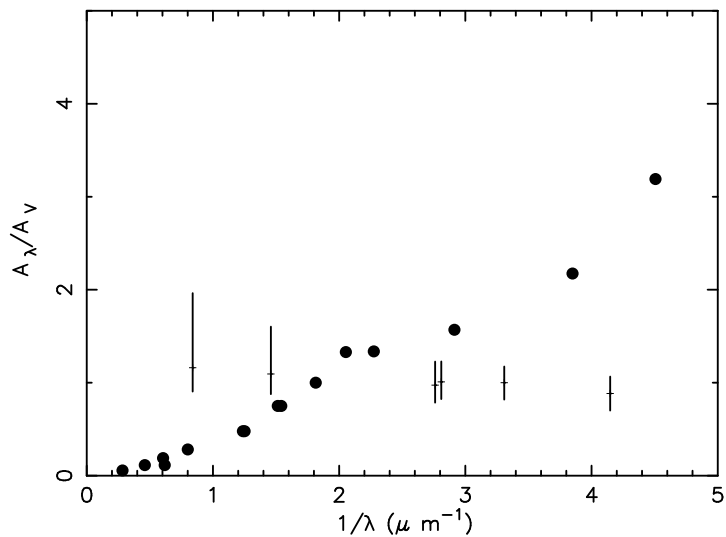} 
  \end{center}
  \caption{Extinction curves (normalized at the $V$ band)
for the host galaxy of GRB~080506 estimated with a simple power-law 
SED extrapolated from the X-ray region.
The bar and filled circles show the curves for GRB~080506 
and MW.
}
\label{fig:extin08}
\end{figure*}

These epochs correspond to the phases defined in table~4, 
 that is, (a) the prompt emission phase ($T+145$ ---$T+200$~s), 
 (b) the steep decay phase ($T+200$ --- $T+350$~s), 
 (c) the flare rise phase ($T+350$ --- $T+474$~s),
 (d) the flare peak phase ($T+474$ --- $T+503$~s), 
 (e) the flare decay phase ($T+503$ --- $T+672$~s), and 
 (f) the normal decay phase ($T+10900$ --- $T+70300$~s). 
As can be seen in the figure, the spectral slope of X-rays 
 clearly changed, while no significant change was detected in 
 the optical---NIR range during the X-ray flare phase from 
 panel (c) to (e).

As well as the case of GRB~071112C, the NIR--UV fluxes are much 
lower than those expected from the X-ray spectra indicated by 
the solid line in the figure.  Thus, the observed NIR--UV fluxes 
were definitely reduced by the interstellar extinction in 
the host galaxy also in the case of GRB~080506.
Since the lack of a measured redshift prevents us from determining
$N_H$ based on the X-ray spectrum, 
we cannot apply the same approach for the absorption correction
 to GRB~080506, as in GRB~071112C.

In the framework of the external shock model, 
the condition of $\alpha_{\rm opt}=\alpha_{\rm X}$ means 
that $\nu_c$ is below the optical frequency, and hence, 
an optical---X-ray SED should be described with a single power-law, namely 
$\beta_{\rm opt}=\beta_{\rm X}$.
In the case of GRB~080506, the observed $\alpha_{\rm opt}$ was in 
agreement with $\alpha_{\rm X}$ within errors after $T+1.5$~ks.  
Hence, the standard external shock model predicts 
$\beta_{\rm opt}=\beta_{\rm X}$ in GRB~080506 after $T+1.5$~ks.
We calculated $\beta_{\rm opt}=0.95 \pm 0.05$ 
shown by dotted line in panel (f) of figure~\ref{fig:sed08},
using the NIR--UV data after $T+1.5$~ks in which
the galactic extinction was corrected \citep{sch98extinct}. 
This is, however, significantly different from $\beta_{\rm X}$ 
 in the normal decay phase, as shown in table~\ref{tab:grb080506_x}.
This is possibly due to a significant reddening 
of the optical afterglow 
in the host galaxy of the GRB.

Then, we can estimate the extinction in the host galaxy in the 
NIR---UV range, 
assuming a single power-law SED between the NIR---X-ray range.
We defined the extinction, $A_\lambda$, 
as $A_\nu = F_\lambda^0/F_\lambda^{\rm obs}$, 
where $F_\lambda^0$ is the flux extrapolated from the X-ray power-law 
spectrum, $F_\nu^0 = F_{\rm X} (\nu/\nu_{\rm X})^\beta$
where $F_{\rm X}$ the observed X-ray band($\nu_{\rm X}$) flux.
$F_\lambda^{\rm obs}$ is the observed NIR--UV fluxes.
Figure~\ref{fig:extin08} shows the 
obtained extinction curve.  The extinction $A_\lambda$ was normalized 
at $A_V$.
The extinction curves in the MW is also 
shown for comparison in figure~\ref{fig:extin08} \citep{car89extinct}.  
In general, the extinction is larger
in the UV region rather than the NIR one.
In the case of GRB~080506, however, the extinction 
is larger in the NIR than in the UV region.
This result is problematic for the standard extinction model with 
 dusts \citep{car89extinct}.  Thus, the extraordinary extinction 
 curve in figure~\ref{fig:extin08} indicates 
 $\beta_{\rm opt}\neq \beta_{\rm X}$ in the normal decay phase 
 of GRB~080506, while $\alpha_{\rm opt}=\alpha_{\rm X}$.  
In the case that we correct the flux with a general extinction 
 law, the SED definitely 
 has a break between the NIR and X-ray bands since the observed 
 $\beta_{\rm opt}$ is smaller than $\beta_{\rm X}$.  As well as the 
 $\alpha_{\rm X}$--$\beta_{\rm X}$ relation reported in \S~3.3, the 
 condition in GRB~080506 is inconsistent with the standard external 
 shock model.

\section{Discussion}     
\begin{table*}
 \caption{$\nu_c$ estimated for possible cases on $\nu_m^\prime$ and 
$\nu_c^\prime$ in the two-component external shock model.
The checks ($\surd$) show the cases that 
the present model can reproduce the 
observed SED variations.}
\label{tab:results_2jets}
\centering          
 \begin{tabular}{lllcc}
\hline\hline
 case & condition  &         \multicolumn{3}{c}{$\nu_c$~(Hz)}   \\ \cline{3-5} 
     &            & formula & GRB~071112C                      & GRB~080506   \\ 
\hline
S1& $\nu_m^\prime < \nu_{\rm O} < \nu_c^\prime < \nu_{\rm X}$&
$\left( \frac{a}{A}\right)^2  10^{3(p-p^\prime)} \nu_c^\prime$   
 & $< 1\times 10^{14}$   & $<1 \times 10^{10}$        \\

S2& $\nu_m^\prime < \nu_{\rm O} < \nu_{\rm X} < \nu_c^\prime$&
$\left( \frac{a}{A}\right)^2  10^{3(p-p^\prime)+18} $
 & $< 6 \times 10^{14}$ $\surd$  & $< 5 \times 10^{11}$    \\

S3& $\nu_{\rm O} < \nu_m^\prime < \nu_c^\prime < \nu_{\rm X}$&
$\left( \frac{a}{A}\right)^2  10^{3p-18p^\prime+5}  \nu_m^{\prime p^\prime -1} \nu_c^{\prime 5/3}$
 & see figure~\ref{fig:f3s3} $\surd$ & see figure~\ref{fig:f3s3}     \\

S4 &$\nu_{\rm O} < \nu_m^\prime < \nu_{\rm X}< \nu_c^\prime $&
$\left( \frac{a}{A}\right)^2  10^{3p-18p^\prime+23} \nu_c^\prime$
 & $< 3 \times 10^{17}$ $\surd$ & $<3 \times 10^{15}$  $\surd$     \\

S5& $\nu_m^\prime < \nu_c^\prime < \nu_{\rm O} < \nu_{\rm X}$&
$\left( \frac{a}{A}\right)^2  10^{33(p-p^\prime) +15}$
 & $<3 \times 10^{-8}$     & $< 2 $ \\

F1& $\nu_c^\prime < \nu_{\rm O} < \nu_m^\prime < \nu_{\rm X}$&
$\left(\frac{a}{A}\right)^2  10^{3p-9p^\prime+30}\nu_c^{\prime (p^\prime -1)}$
 &$< 1\times 10^{22}$ $\surd$ &  $<5 \times 10^{36}$ $\surd$ \\ 

F2& $\nu_c^\prime < \nu_{\rm O} < \nu_{\rm X} < \nu_m^\prime$&
$\left( \frac{a}{A}\right)^2  10^{3p+12}$
 &$< 2 \times 10^{13}$   & $< 1 \times 10^{15}$ $\surd$\\

F3& $\nu_{\rm O} < \nu_c^\prime < \nu_m^\prime < \nu_{\rm X}$&
$\left( \frac{a}{A}\right)^2  10^{3p-18p^\prime+5}  \nu_m^{\prime p^\prime -1/3} \nu_c^\prime$
 & see figure~\ref{fig:f3s3} $\surd$ & see figure~\ref{fig:f3s3}     \\

F4& $\nu_{\rm O} < \nu_c^\prime < \nu_{\rm X}< \nu_m^\prime $&
$\left(\frac{a}{A}\right)^2  10^{3p-18p^\prime+5}\nu_c^{\prime p^\prime +2/3}$   
 &$< 5 \times 10^{16}$ $\surd$ & $< 6 \times 10^{13}$  \\

F5& $\nu_c^\prime < \nu_m^\prime < \nu_{\rm O} < \nu_{\rm X}$&
$\left( \frac{a}{A}\right)^2  10^{33(p-p^\prime) +15} \nu_m^\prime$   
 &$< 3 \times 10^{-8}$   & $< 4 \times 10^{13}$ \\
\hline
\end{tabular}
\end{table*}

\begin{figure*}
\centering
   \includegraphics[width=140mm]{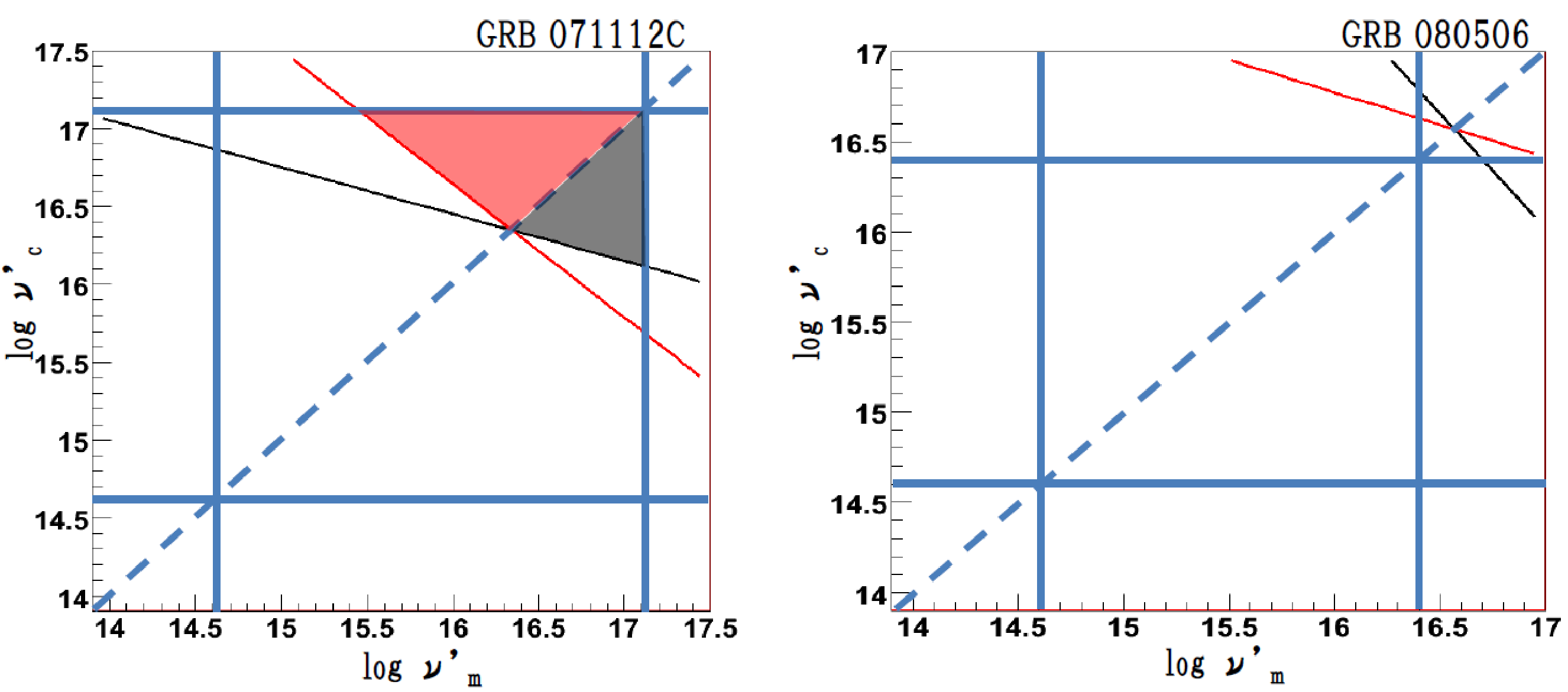}
\caption{
Allowed region of $\nu_c^\prime$ and $\nu_m^\prime$.
The red and gray shaded regions show the areas 
in case F3 and S3, respectively.
The blue lines show the assumed condition for $\nu_c^\prime$ 
and $\nu_m^\prime$.
These frequencies are $6\times10^{14}~{\rm Hz}$--$2\times10^{17}$~Hz
and $5\times10^{14}~{\rm Hz}$--$2\times10^{16}$~Hz
for GRB~071112C and GRB~080506, respectively.
Left and right panels show the cases of GRB~071112C and GRB~080506, 
respectively.
}
\label{fig:f3s3}
\end{figure*}

In this section, we discuss 
whether the X-ray flare originates in the external or internal shocks.

If they arise from external shocks,
optical and NIR flares should be
contemporaneously detected during the X-ray flares \citep{Fan05}.
In addition, the spectral index in the optical band should also 
change (e.g. \citealt{Zhang06}).
Our observations unambiguously show no optical variation 
associated with the X-ray flares.
The lack of optical variations is consistent with previously 
reported observations 
(e.g. \citealt{Burrows05a, Krimm07, Butler_Kocevski07}).  
The optical spectral index also unchanged during the X-ray flares
in both GRB~071112C and 080506.
Therefore, the optical behavior during these X-ray flares is 
inconsistent with the prediction from the standard external shock 
model.
On the other hand, the
late internal shock model can explain the X-ray flare even if
no contemporaneous flares are observed in the optical and NIR 
bands \citep{Burrows05a}.
In fact, the lack of the NIR--optical variations in both GRB~071112C 
and GRB~080506 is consistent with this model.


In the following,
we propose an alternative model for the X-ray flare.
We consider two components of the external shocks;
one originates in the main shell which produces the normal afterglow,
and the other originates in a delayed shell 
which is responsible for the X-ray flare.
The observed sharp decay of the X-ray flare can be explained by 
this scenario because the decay slope of the emission from 
the delayed shell can be apparently steep in the time frame of 
the normal afterglow (\citealt{Zhang06b, Yamazaki06}).
The delayed shell could generate a prominent X-ray emission under the 
condition that it passes a region containing enough interstellar medium.  
Such a condition is probably achieved, for example, if the opening 
angle of the delayed jet is larger than that of the main jet, 
or if the axis of the delayed jet is off to that of the main jet.
These conditions were originally proposed 
to explain `X-ray flashes', which are analogous to GRBs except for 
their softer emission and less energetics 
\citep{Yamazaki02,Yamazaki03,Yamazaki04,Lamb05,L.Li06}.
\citet{Piro05} and \citet{Galli06} have proposed models 
for the X-ray flare, in which the delayed shell interacts
with the reverse shock region of the preceding jet.

In our model, we assume that the temporal evolution of the emission 
from the delayed shell is the same as that of the main shell, 
following the standard external shock model \citep{sar98grblc}.  
For the decay phase of the X-ray flare,
we assumed that the spectrum of delayed shell overlaps that of the main
shell.
We denote the maximum and cooling frequencies of the synchrotron emission 
as $\nu_m$ and $\nu_c$ for the main shell, and
$\nu_m^\prime$ and $\nu_c^\prime$ for the delayed shell.
Then, we investigated several conditions of $\nu_m^\prime$ and $\nu_c^\prime$,
which satisfy the observed amplitudes of variations and 
electron energy distribution index.

We have mentioned in $\S$ \ref{sec:sed07} and \ref{sec:sed08}
that for both events we analyzed,
the SEDs of the normal afterglow component are 
expected to have a break between the NIR and X-ray bands.  
This break means that $\nu_c$ lies in this bands.
There is no spectral break 
in the observed NIR--optical SEDs and X-ray spectra.  
Therefore, $\nu_c$ are constrained as
$6\times10^{14}~{\rm Hz}<\nu_c<2\times10^{17}$~Hz
and $5\times10^{14}~{\rm Hz}<\nu_c<2\times10^{16}$~Hz
for GRB~071112C and GRB~080506, respectively.  
The above frequencies are those in the observer's frame.  
Because we focus on the ratio of the frequency in the 
following discussion, the redshift correction to the 
energy band is not important.

                   
The cooling frequency of the normal afterglow
$\nu_c$, is also constrained using the observed amplitudes 
of the X-ray flares.
In this procedure, we consider ten cases depending  
on the relationship among $\nu_m^\prime$, $\nu_c^\prime$, 
and the observation frequencies of $\nu_{\rm O}$ and $\nu_{\rm X}$.
As listed in table~\ref{tab:results_2jets},
possible conditions are divided 
into five cases in the slow cooling regime ($\nu_m^\prime < \nu_c^\prime$) 
and  five cases in the fast cooling regime ($\nu_c^\prime < \nu_m^\prime$).
Since we discuss the SED of the decay phase of the X-ray flare,
the conditions are limited to those ten cases;
the conditions of $\nu_m^\prime < \nu_X$ and 
$\nu_c^\prime < \nu_X$ are required for the decay phase in 
the slow and fast cooling regime, respectively.
The cases for the slow and fast cooling regimes are indicated 
by the characters `S' and `F', respectively.

\begin{table}
 \begin{center}
 \caption{Observational parameters used in the two-component 
 external shock model.}
 \label{tab:Aap}
  \begin{tabular}{lll}
     \hline \hline
results & GRB~071112C           & GRB~080506   \\ \hline
 $A$   & 3.5                    & 15         \\
 $a$   & $<$ 0.07               & $<$ 0.14       \\
 $p$   & 1.66  ($\pm$0.10)      & 2.70 ($\pm$0.34)       \\
 $p^\prime$  & 1.24 ($\pm$0.26) & 2.92  ($\pm$0.20)     \\
\hline
   \end{tabular}
 \end{center}
\end{table}


For example, in case~S1 in table~\ref{tab:results_2jets},
the X-ray and optical flux densities of the normal decay component, 
$F_{\rm n, X}$ and  $F_{\rm n, O}$ are related as;
\begin{eqnarray}
\label{eq:fnx}
F_{\rm n, X} = F_{\rm n, O}
      \left( \frac{\nu_c}{\nu_{\rm O}}\right)^{-\frac{p-1}{2}}
      \left( \frac{\nu_{\rm X}}{\nu_c}\right)^{-\frac{p}{2}}.
\end{eqnarray}
Here we assumed an SED predicted by the standard external shock 
model \citep{sar98grblc}.
Similarly, 
the optical and X-ray flux densities of the X-ray flare component, 
$F_{\rm flare, O}$ and $F_{\rm flare, X}$ are related as;
\begin{eqnarray}
\label{eq:s1ffo}
F_{\rm flare, O} = F_{\rm flare, X}
      \left( \frac{\nu_c^\prime}{\nu_{\rm X}}\right)^{-\frac{p^\prime}{2}}
      \left( \frac{\nu_{\rm O}}{\nu_c^\prime}\right)^{-\frac{p^\prime-1}{2}},
\end{eqnarray}
where $p^\prime$ is the electron distribution 
index of the X-ray flare component.
We define the ratios of the flare to the normal fluxes in the
X-ray and optical bands as 
\begin{eqnarray}
\label{eq:fx}
A = F_{\rm flare, X} / F_{\rm n, X},
\end{eqnarray}
\begin{eqnarray}
\label{eq:fo}
a = F_{\rm flare, O} / F_{\rm n, O}.
\end{eqnarray}
Using Eqs. (\ref{eq:fnx}), (\ref{eq:s1ffo}), (\ref{eq:fx}), 
and (\ref{eq:fo}), we obtain
\begin{eqnarray}
\label{eq:uehara_eq}
    \nu_c = \left(  \frac{a}{A}\right)^2  10^{3(p-p^\prime)} \nu_c^\prime~~,
\end{eqnarray}
where we  take $\nu_{\rm X}/\nu_{\rm O} \approx 10^{3}$. 
In this case S1, the cooling frequency of the flare component, $\nu'_c$,
is assumed to lie between $\nu_{\rm O}$ and $\nu_{\rm X}$.
Given this fact, and substituting observed quantities summarized in
 table~\ref{tab:Aap} into Eq.~(\ref{eq:uehara_eq}),
we derive the condition for the cooling frequency of the normal afterglow
component as $\nu_c < 1 \times 10^{14}$ and  
$< 4 \times 10^{10}$~Hz  for GRB~071112C and GRB~080506, respectively.
Those $\nu_c$ was calculated taking into account the
uncertainties of $p$ and $p\prime$ in order to obtain
firm estimates.
Both for GRB~071112C and 080506, two independent conditions
give no allowed regions for $\nu_c$, so that case S1 fails
to explain the observed results.


Similar to case~S1,
$\nu_c$ in the other nine cases are also evaluated.
Table~\ref{tab:results_2jets} shows the results.
Upper limits of $\nu_c$ are given in the table 
if it is independent of $\nu_c^\prime$ and $\nu_m^\prime$ 
or if it is a function of either of them
(cases S1, S2, S4, F1, F2, and F4).  
If $\nu_c$ depends 
on both $\nu_c^\prime$ and $\nu_m^\prime$ (cases S3 and F3),
then in the $\nu_c^\prime$--$\nu_m^\prime$ plane
we search for allowed regions to satisfy the conditions of 
$6\times10^{14}~{\rm Hz}<\nu_c<2\times10^{17}$~Hz
and $5\times10^{14}~{\rm Hz}<\nu_c<2\times10^{16}$~Hz
for GRB~071112C and GRB~080506, respectively.  
  Figure~\ref{fig:f3s3} shows the results.  
The left and right panels of Figure~\ref{fig:f3s3} are 
for GRB~071112C and GRB~080506, respectively.  
The blue solid and dotted lines indicate the assumed conditions 
for $\nu_c^\prime$ and $\nu_m^\prime$. 
The red and gray shaded regions indicate the allowed region 
for $\nu_c$ in cases F3 and S3, respectively.
Thus, both cases F3 and S3 can explain the observation with 
the two-component external shock model for only GRB~071112C. 
On the other hand, there is no allowed region in the
 case of GRB 080506, as can be seen in the right panel
 of Figure~\ref{fig:f3s3}.
The `check' symbols are given in table~\ref{tab:results_2jets} 
in the case that there are allowed values of parameters 
that can reproduce the observed SED variations.
For both GRBs, there are several cases in which the observations
can be explained by our model.


Note that all parameters needed for the above discussion are 
the variation amplitudes during the X-ray flares in the optical and 
X-ray ranges, $a$ and $A$, 
and electron energy distribution index, $p$ and $p^\prime$.
Thus, the discussion was independent of the uncertainty in
the correction of the dust extinction in the GRB host galaxies.

\section{Conclusion}
We have observed GRB~071112C and GRB~080506 
before the start of X-ray flares.
In conjunction with published X-ray and optical data, 
we analyzed densely sampled light curves of the early afterglows 
and SEDs of the NIR--X-ray range.
We found that the SEDs in the normal decay phase had a break between 
the UV and soft X-ray regions.
No significant variation in the optical-NIR range was detected contemporaneous to the
X-ray flares.
The lack of the optical--NIR variation suggests that the late 
internal shock is a reasonable origin for the X-ray flare.  
In addition, we found that two-component external shock model can 
also explain the observed variations of SEDs during the X-ray flares.

\vspace{1cm}
The authors thank the referee for careful reading and many useful comments.
The authors also thank the $Swift$ team 
for development of hardware/software
and operation. SN is supported by Research Fellowships
of the Japan Society for the Promotion of Science for
Young Scientists.


\end{document}